\documentclass[a4paper]{ISOStyle}
\usepackage{epsfig}

\begin{document}

\title{Gas-phase H$_2$O and CO$_2$ toward massive protostars}

\author{A.M.S. Boonman\inst{1} \and E.F. van Dishoeck\inst{1} \and 
F. Lahuis\inst{2,3} \and C.M. Wright\inst{4} \and S.\,D. Doty\inst{5}} 
\institute{Leiden Observatory, P.O. Box 9513, NL--2300 RA Leiden, The
  Netherlands
  \and Space Research Organization Netherlands, P.O. Box 800, 9700 AV, 
  Groningen, The Netherlands
  \and ISO Science Operations Centre, Astrophysics Division, Space Science 
  Department of ESA, Villafranca, P.O. Box 50727, E-28080 Madrid, Spain
  \and School of Physics, University College, ADFA, UNSW, Canberra ACT 
  2600, Australia 
  \and Department of Physics and Astronomy, Denison University, Granville, 
  Ohio 43023, USA}

\maketitle 

\begin{abstract}
We present a study of gas-phase H$_2$O and CO$_2$ toward a sample of 14 
massive protostars with the ISO-SWS. Modeling of the H$_2$O spectra using a 
homogeneous model with a constant excitation temperature $T_{\rm ex}$ shows 
that the H$_2$O abundances increase with temperature, up to a few times 
10$^{-5}$ with respect to H$_2$ for the hottest sources 
($T_{\rm ex}\sim$ 500~K). This is still a 
factor of 10 lower than the H$_2$O ice abundances observed toward cold 
sources in which evaporation is not significant (Keane et al.\ 2001). 
Gas-phase CO$_2$ is not abundant in our sources. The abundances are 
nearly constant for $T_{\rm ex} \ga$ 100 K at a value of a
few times 10$^{-7}$, much lower than the solid-state abundances of
$\sim$1--3$\times 10^{-6}$ (Gerakines et al.\ 1999). 
For both H$_2$O and CO$_2$ the gas/solid ratio increases 
with temperature, but the increase is much stronger for H$_2$O than for CO$_2$,
suggesting a different type of chemistry. In addition to the homogeneous 
models, a power law model has been developed for one of our sources,
based on the physical structure of this region as determined from 
submillimeter data by van der Tak et al. (1999).
The resulting H$_2$O model spectrum gives a good fit to the data.  
\end{abstract}

\section{INTRODUCTION}

The Infrared Space Observatory (ISO) has provided us with 
a wealth of new data in the infrared from regions of massive star formation.
This includes unique information on molecules such as H$_2$O and CO$_2$, 
which are difficult to observe from the ground due to the Earth's atmosphere.
H$_2$O and CO$_2$ are among the most abundant species in the envelopes of 
massive protostars and play a key role in the chemistry in these 
regions. H$_2$O is a particularly powerful molecule to study the interaction of
the protostar with its environment. In warm regions and shocks all gas-phase 
oxygen not locked up in CO is thought to be driven into H$_2$O, predicting 
greatly enhanced gas-phase H$_2$O abundances. Also, its level populations are 
influenced by mid- and far-infrared radiation from warm dust, in addition 
to collisions. Observations of the well-studied star-forming region Orion-KL 
indeed show strong gas-phase H$_2$O lines, corresponding to abundances up to 
10$^{-4}$ (van Dishoeck et al.~1998, Gonzalez-Alfonso et al.~1998, Harwit et 
al.~1998, Wright et al.~2000). High resolution Fabry-P\'{e}rot observations 
of pure rotational 
H$_2$O lines suggest that this gas is associated with warm and shocked regions 
(Wright et al.~2000).
The CO$_2$ molecule on the other hand cannot be observed through rotational
transitions, because it does not have a permanent dipole moment. The
excitation temperature of these types of molecules can be a useful indicator 
of the kinetic temperature of the region.

\hspace*{-0.5cm}We have studied gas-phase H$_2$O and CO$_2$ toward a 
sample of 14 massive young stellar objects including GL~2136, GL~2591, 
W~3 IRS5, NGC~7538 IRS9, MonR2 IRS3 and GL~490 (see Table 1). 
The luminosities of these objects are $\sim$10$^4$--10$^5$ L$_{\odot}$, 
the masses of the envelopes are $\sim$100 M$_{\odot}$ and their distances are 
$\sim$1-4 kpc. Most of these objects show a multitude of 
gas-phase H$_2$O absorption lines around 6 $\mu$m in the SWS spectra,
originating in the $\nu_2$ ro-vibrational band.
 The ro-vibrational band at 15 $\mu$m of gas-phase CO$_2$ has also
been detected in many sources. 
The LWS spectra, however, do not show strong lines of gas-phase H$_2$O 
(Wright et al. 1997). A subset of these sources has been studied previously
by van Dishoeck \& Helmich (1996) and van Dishoeck (1998).

Both molecules have also been detected in the solid phase toward many 
massive protostars (e.g. Gerakines et al.~1999, Keane et al.~2001). 
This allows us to determine gas/solid ratios. 
In addition, most of these sources show a rich submillimeter emission spectrum,
allowing the derivation of temperature and density profiles 
(van der Tak et al.~2000).
\begin{figure}[ht]
\vspace*{-0.5cm}
  \begin{center}
    \hspace*{-0.5cm}\psfig{file=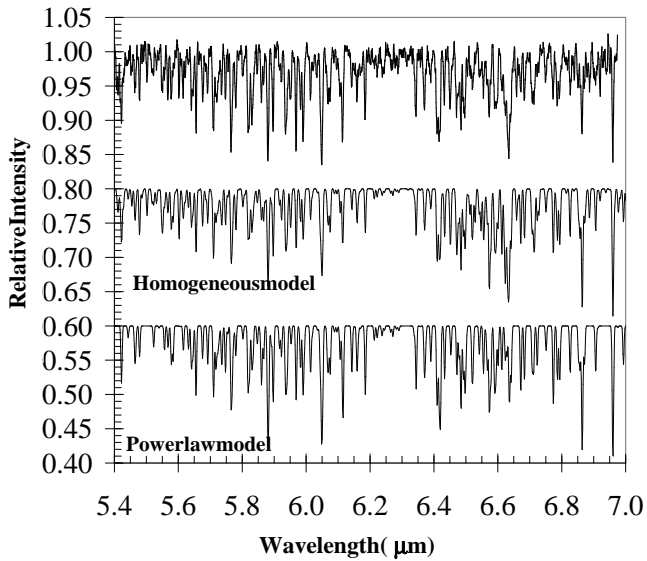, width=9.0cm}
  \end{center}
%\resizebox{\hsize}{!}{\includegraphics{ISOpost_tot.eps}}
\vspace*{-5cm}
\caption{H$_2$O spectra toward the massive protostar GL~2591. The top panel 
shows the data, the middle and bottom panel the homogeneous and power law 
model, respectively. \label{SPU_fig1}}
\end{figure}

\section{OBSERVATIONS AND REDUCTION}

The $\nu_2$ ro-vibrational bands of gas-phase H$_2$O and CO$_2$ around 
6 $\mu$m and 15 $\mu$m, respectively, have been observed with the Short 
Wavelength Spectrometer (SWS) in the AOT6 grating mode. All spectra have 
been reduced using the standard pipeline reduction 
routines starting from SPD level. In addition, the instrumental fringes have
been removed by fitting a cosine to the data (Lahuis \& van Dishoeck~2000).
The 6 $\mu$m spectra have been rebinned to a spectral resolution of 
$\Delta$$\lambda$=0.0020 $\mu$m and the 15 $\mu$m spectra to 
$\Delta$$\lambda$=0.0035 $\mu$m. The $S/N$ ratio on the continuum  
is typically 50-100 in the final spectra. 
\begin{figure}[ht]
\vspace*{-0.8cm}
  \begin{center}
    \psfig{file=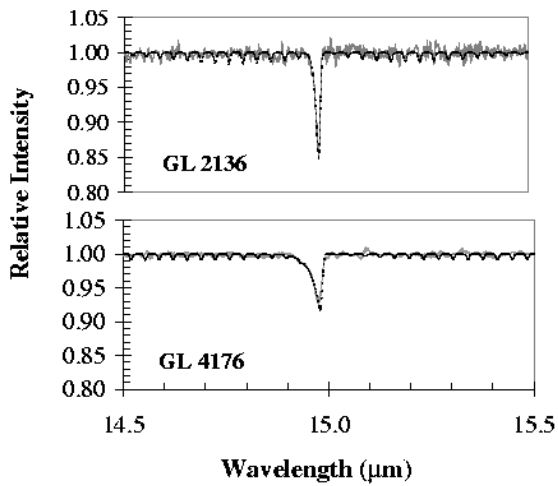, width=7cm, angle=270}
  \end{center}
%\resizebox{\hsize}{!}{\includegraphics{ISOpost_tot.eps}}
%\vspace*{-7cm}
%\vspace*{-7cm}
\caption{CO$_2$ spectra toward the massive protostars GL~2136 and GL~4176. 
The grey solid lines are the data, the black dashed lines the models. The 
solid CO$_2$ feature has been divided out.
\label{SPU_fig2}}
\end{figure}

\begin{table}[ht]
  \caption{Preliminary results from modeling of the H$_2$O and CO$_2$ 
spectra using constant $T_{\rm ex}$.}
\label{SPU_tab:table}
\begin{center}
\leavevmode
\footnotesize{
\begin{tabular}{lcccc}\\ [-20pt]
\hline
\noalign{\smallskip}
Source&$T_{\rm ex}$(H$_2$O)&$N$(H$_2$O)$^e$&$T_{\rm ex}$(CO$_2$)&$N$(CO$_2$)\\
&K&10$^{18}$cm$^{-2}$&K&10$^{16}$cm$^{-2}$\\
\noalign{\smallskip}
\hline
\noalign{\smallskip}
GL2136&500&1.5&175&2.7\\ 
GL2591&450&3.5&500&2.5\\
GL4176&400&1.5&500&2.5\\
GL2059&500&1.0&500&$<$0.3\\
MonR2 IRS3&300&0.6&300&2.0\\
GL490&107$^a$&$<$0.3&107&$<$0.2\\
NGC3576&500&1.5&500&$<$0.7\\
NGC7538 IRS1&176$^b$&$<$0.2&400&0.8\\
NGC7538 IRS9&180$^b$&$<$0.2&150&0.8\\
NGC2024 IRS2&44$^c$&$<$0.09&44&$<$0.2\\
S140 IRS1&390$^b$&$<$0.2&390&$<$0.6\\
W33A&120$^b$&$<$0.2&300&2.3\\
W3 IRS4&55$^d$&$<$0.3&80&$<$0.3\\
W3 IRS5&400&0.4&350&0.7\\
\noalign{\smallskip}
\hline
\noalign{\smallskip}
\end{tabular}
}
\end{center}
{\footnotesize
$^a$ $T_{\rm ex}$(CO) from Mitchell et al. (1995)\\
$^b$ $T_{\rm ex}$($^{13}$CO) from Mitchell et al. (1990)\\
$^c$ $T_{\rm ex}$($^{13}$CO) from Black \& Willner (1984)\\
$^d$ $T_{\rm kin}$ from Helmich (1996)\\
$^e$ Assuming $b$=5 km s$^{-1}$\\
}
\end{table}

\section{MODELING}

The modeling of the spectra has been performed using synthetic spectra from 
Helmich (1996), assuming a homogeneous source with a single temperature 
$T_{\rm {ex}}$ and column density $N$. Since the H$_2$O models are 
sensitive to different Doppler $b$-values, a range of values between 1.5 and 
10 km s$^{-1}$ has been used. For CO$_2$ the models are not sensitive to the 
linewidth, so a mean value of $b$=3 km s$^{-1}$ is adopted here. This is in 
agreement with observations of other ro-vibrational absorption lines in the 
same wavelength region toward these sources. The best fit to the data has 
then been 
determined using the reduced $\chi_{\nu}^2$-method. Some good fitting models 
are shown in Fig. 1 and 2 for H$_2$O and CO$_2$ respectively. An example of 
$\chi_{\nu}^2$ contours is shown in Fig. 3 for the source GL~2136. This figure
illustrates that for low $b$-values (i.e. $b<$2.5 km s$^{-1}$) the 
temperature and column density of the gas-phase H$_2$O is not well 
constrained. In the following analysis $b$=5 km s$^{-1}$ is adopted for H$_2$O.
This corresponds to the mean value of the $^{13}$CO $v$=1-0 absorption line 
widths found by Mitchell et al. (1990). 
%\vspace*{-1.5cm} 
\begin{figure}[ht]
\vspace*{-1.5cm} 
  \begin{center}
    \psfig{file=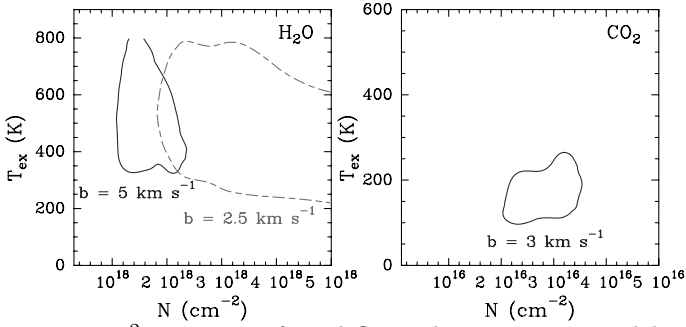, width=6.5cm, angle=270}
  \end{center}
%\resizebox{\hsize}{!}{\includegraphics{ISOpost_tot.eps}}
\vspace*{-1.5cm} 
\caption{$\chi_{\nu}^2$-contours of good fitting homogeneous models for 
H$_2$O and CO$_2$ toward the massive protostar GL~2136. \label{SPU_fig3}}
\end{figure}

\section{ANALYSIS}
\subsection{H$_2$O}
\subsubsection{Homogeneous models}

The results of the homogeneous model analysis for H$_2$O are presented in 
Fig. 4 and 
Table 1. It is seen that sources with higher 
excitation temperatures have higher column densities, 
although the increase is not very strong. The column densities have been 
converted into abundances with respect to the hot H$_2$ gas, since the 
spectra show primarily the warmer H$_2$O gas. The H$_2$ column densities have 
been  
derived from infrared observations of $^{13}$CO assuming a $^{12}$CO/$^{13}$CO
ratio of 60 and a $^{12}$CO/H$_2$ ratio of 2$\times$10$^{-4}$ (e.g. Mitchell 
et al.~1990, Lacy et al.~1994). The resulting abundances increase 
with temperature, up to a few times 10$^{-5}$ for the 
hottest sources ($T_{\rm ex}\sim$ 500 K) (Fig.4).
The presence of strong C$_2$H$_2$ absorption toward the same sources
(Lahuis \& van Dishoeck~2000) and the absence of this molecule in the 
well-known shocked regions Peak 1 and Peak 2 in Orion (Boonman et al. 2001) 
suggest that shocks do not play a dominant role. 
Hot core models by Charnley (1997) indicate H$_2$O abundances of 
$\sim 10^{-5}$ for 
$T$=300 K, consistent with our values. However, he assumes that the 
initial solid-state abundance of H$_2$O is $\sim$10$^{-5}$, a factor of 10 
lower than observed toward cold sources in which evaporation is not 
significant (Keane et al.\ 2001). This discrepancy suggests that part of 
the evaporated H$_2$O is probably destroyed through rapid gas-phase reactions 
leading to atomic oxygen.
%\vspace*{-1.5cm}
%\verb!\begin{figure*}![ht]
\begin{figure}[ht]
\vspace*{-0.5cm}
  \begin{center}
    \psfig{file=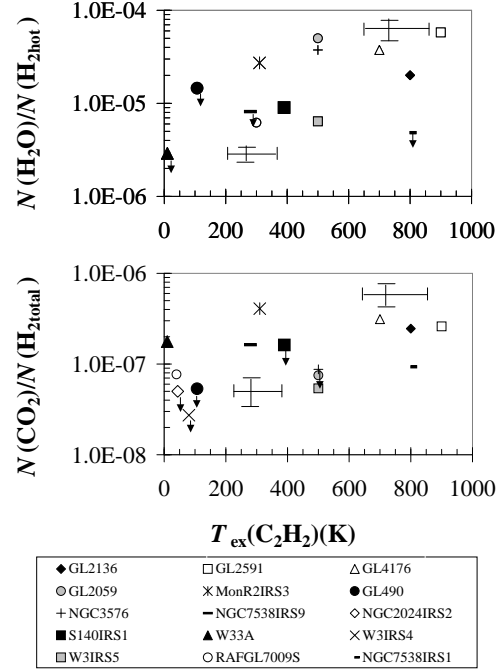, width=7.5cm}
  \end{center}
%\resizebox{\hsize}{!}{\includegraphics{ISOpost_tot.eps}}
\vspace*{-1.8cm}
\caption{{\it Top panel:} H$_2$O abundances with respect to hot H$_2$ gas. 
{\it Bottom panel:} CO$_2$ abundances with respect to total H$_2$ gas.
The excitation temperature of C$_2$H$_2$ is used as a tracer of the warm gas 
(see Lahuis \& van Dishoeck~2000). Typical errorbars are indicated in the 
figure. The intermediate mass protostar AFGL~7009S 
(Dartois et al.~1998) is added for comparison.\label{SPU_fig4}}
\end{figure}
%\verb!\end{figure*}!

\subsubsection{Power law model}

Although the homogeneous models provide a good fit to the data, they probably
do not reflect the true excitation mechanism of the H$_2$O molecule. 
Since the level populations of H$_2$O are influenced by 
radiation from warm dust, pumping has to be included in the models. 
Also submillimeter observations show that both a temperature and density 
gradient is present in these objects (van der Tak et al. 2000), and therefore 
an abundance gradient. Therefore 
we have set up a power law  model for one of our sources GL~2591, using the
models by Doty \& Neufeld (1997). In this model a density gradient 
$\propto$ r$^{-1.25}$ (van der Tak et al. 1999) is used. The temperature and 
abundance profiles are shown in Fig. 5. Only gas-phase chemistry is included 
at this point. Although Doty \& Neufeld report no significant changes in 
their models if gas-grain chemistry is included, further 
investigations have to confirm this. Similar models 
will be set up for all other sources in our sample. 
The resulting model spectrum for GL~2591 (Fig. 1) gives a good fit to 
the data. 
\begin{figure}[ht]
\vspace*{-0.7cm}
  \begin{center}
    \psfig{file=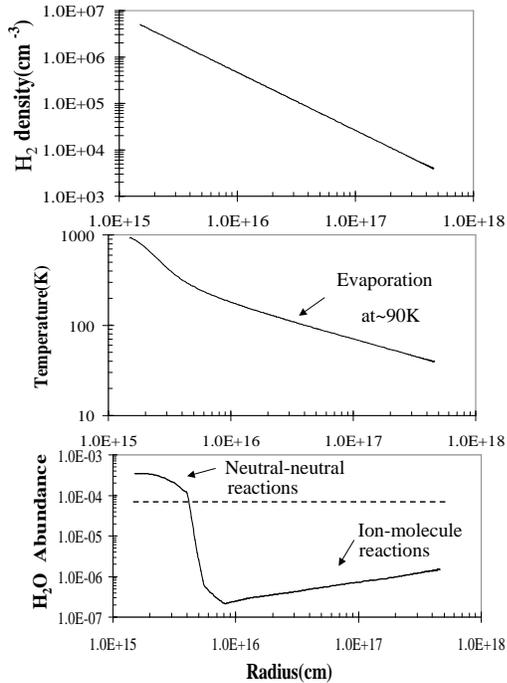, width=7.5cm}
  \end{center}
%\resizebox{\hsize}{!}{\includegraphics{ISOpost_tot.eps}}
\vspace*{-1cm}
\caption{Power law models for H$_2$O based on the model by Doty \& Neufeld 
(1997) for GL 2591. The dashed line denotes the H$_2$O abundance found from 
the homogeneous models. \label{SPU_fig5}}
\end{figure}
\vskip 10pt

\hspace*{-0.5cm}Because of the high H$_2$O abundance derived from the 
ro-vibrational lines, 
some pure rotational H$_2$O lines are expected in the LWS spectra. However, 
reduction of these spectra for GL~2591 only shows a hint of one 
line in absorption (Wright et al. 1997). A first quick look at the 
LWS spectra of the other sources shows only a few pure 
rotational lines of H$_2$O. 
Since the LWS beam is so large ($\sim$80$^{\prime\prime}$ diameter) compared 
to the angular size of the infrared sources, this suggests that the hot 
water fills only a small fraction of the LWS beam ($<$ few arcseconds) 
close to the protostar, consistent with Fig. 5.

\subsection{CO$_2$}

The homogeneous model analysis for gas-phase CO$_2$ shows that this molecule 
is not very abundant in our sources. The column densities show only a weak 
increase with temperature, whereas the abundances are roughly constant for 
$T_{\rm ex} \ga$ 100 K at a value of a few $\times$10$^{-7}$. Since both 
warm and cold CO$_2$ is detected, the abundances are given with 
respect to the total H$_2$ column density (Fig. 4).
These abundances are much lower than the solid-state abundances of
$\sim$1--3$\times 10^{-6}$ (Gerakines et al.\ 1999). This suggests that CO$_2$
is also being rapidly destroyed in the gas-phase after evaporation from the 
grains. Shock chemistry has been suggested by Charnley \& Kaufman (2000), but 
more detailed models including evaporation have to be developed to 
determine the nature of these reactions.

\subsection{Gas/solid ratios}

From the column densities derived from the homogeneous models for H$_2$O and 
CO$_2$, gas/solid ratios can been determined, using the solid-state features 
of H$_2$O (Keane et al.\ 2001) and CO$_2$ (Gerakines et al.\ 1999) as observed
with ISO-SWS toward the same objects. For both species this ratio increases 
with temperature, consistent with the location of both species in the warm 
inner part of the envelope.
However the increase is much stronger for H$_2$O than for CO$_2$,
although CO$_2$ is more volatile than H$_2$O. The higher ratios for the warmer
sources indicate that they are in a later evolutionary stage than 
the sources with low gas/solid ratios (van der Tak et al. 2000, van Dishoeck 
\& van der Tak 2000).
\begin{figure}[ht]
\vspace*{-1.7cm}
  \begin{center}
    \hspace*{-0.5cm}\psfig{file=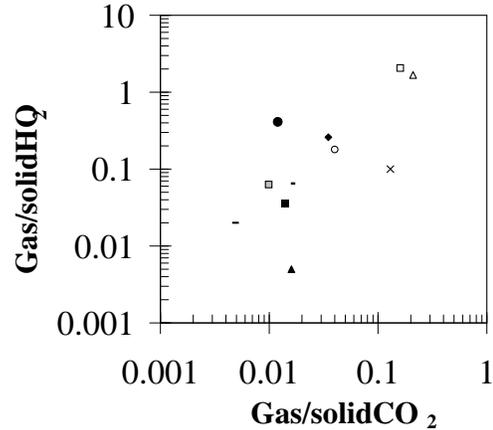, width=8.5cm, angle=270}
  \end{center}
%\resizebox{\hsize}{!}{\includegraphics{ISOpost_tot.eps}}
\vspace*{-1.7cm}
\caption{Gas/solid ratios for H$_2$O and CO$_2$. The symbols are the same as 
in Fig. 4. \label{SPU_fig6}}
\end{figure}

\section{Conclusions}

Gas-phase H$_2$O and CO$_2$ are detected towards a large
number of massive protostars with ISO-SWS. Modeling of the spectra shows that
H$_2$O is hot and abundant. The abundances of gas-phase CO$_2$ are however
not very high. Chemical models will be developed in the near future to 
attempt to explain the differences between these molecules and to investigate 
the possible destruction of these species through gas-phase reactions. 
A power law physical-chemical model for one source shows good agreement
with the data. The LWS data for the same 
sources show mostly a lack of pure rotational H$_2$O lines, indicating that 
the warm gas probed by the ro-vibrational lines is located close to the 
protostar.

\begin{acknowledgements}

The authors are grateful to F. van der Tak, J. Keane and X. Tielens for 
stimulating discussions. This work was supported by the Netherlands 
Organization for Scientific Research (NWO) through grant 614-041-003. 
CMW acknowledges support of an ARC Australian Postdoctoral Fellowship. 
\end{acknowledgements}

\end{document}